\documentclass[twocolumn]{aa}
\usepackage{graphicx}
\usepackage{natbib}

\newcommand{\sax}{{\it Beppo\-SAX}}
\newcommand{\integral}{{\it INTEGRAL}}

\begin{document}

   \title{IBIS preliminary results on Cygnus X-1 spectral and temporal characteristics\thanks{Based
on observations with {\it INTEGRAL},
     an ESA project with instruments and science data centre funded by
     ESA member states (especially the PI countries: Denmark, France,
     Germany, Italy, Switzerland, Spain), Czech Republic and Poland,
     and with the participation of Russia and the USA.}}

   \author{
     A. Bazzano\inst{1},
     A. J. Bird\inst{2},
     F. Capitanio\inst{1},
     M. Del Santo\inst{1},
     P. Ubertini\inst{1},
     A. A. Zdziarski \inst{3},
     G. Di Cocco\inst{4},
     M. Falanga\inst{5},
     P. Goldoni\inst{5},
     A. Goldwurm\inst{5},
     P. Laurent\inst{5},
     F. Lebrun\inst{5},
     G. Malaguti\inst{4},
     A. Segreto\inst{6}
   }

  \offprints{A. Bazzano: angela@rm.iasf.cnr.it}

   \institute{
     Istituto di Astrofisica Spaziale e Fisica Cosmica IASF, Roma, Italy
     \and
     Departement of Physics and Astronomy, Southampton University, Southampton, U.K.
     \and
     Centrum Astronomiczne im.\ M. Kopernika, Bartycka 18, 00-716 Warszawa, Poland
     \and
     Istituto di Astrofisica Spaziale e Fisica Cosmica IASF, Bologna, Italy
     \and
     Service D'Astrophysique, CEA-Saclay, F-91191 Gif-sur-Yvette Cedex, France
     \and
     Istituto di Astrofisica Spaziale e Fisica Cosmica IASF, Palermo, Italy
     }

   \date{Received, accepted}

\abstract{ We report preliminary results of a broadband spectral
and temporal study of the black-hole binary Cyg X-1 performed with
the IBIS telescope. Cyg X-1 was the first pointed celestial target
of IBIS during the {\it INTEGRAL\/} Performance and Verification
Phase, 2002 Nov.--Dec.,  for a total observing time of $\sim$2 Ms
in both staring and dithering mode. Here, we report on only the
staring, on-axis, observation performed in a stable instrument
configuration. During the observing period the source was in its
characteristic low/hard state, in which a few flares and dips have
been detected. The IBIS/ISGRI results demonstrate that the {\it
INTEGRAL\/} observatory offers a unique capability for studying
correlations between hardness and/or flux in different bands over
a wide photon energy range. One of our new results is finding that
the hardness-flux correlation changes the sign twice over the
20--220 keV; first from positive to negative at $\sim$50 keV, and
then back to positive at $\sim$120 keV. The former change appears
to be due to the spectral curvature introduced by variable Compton
reflection. The latter may be due spectral pivoting.
   }

\authorrunning{A. Bazzano et al.}
\titlerunning{IBIS preliminary results on Cygnus X-1}

\maketitle

\section{Introduction}
\label{intro}

Cyg X-1 was the first established black-hole binary (BHB; Webster
\& Murdin 1972; Bolton 1972), and it is so far one of 18 known
BHBs and one of three persistent ones. In the X-ray band, BHBs
usually exhibit five distinct spectral /temporal states: low/hard
(LH), high/soft (HS), very high, intermediate and quiescent (see a
recent review by McClintock \& Remillard 2003). Cyg X-1 itself
exhibits strong variability on all time scales ranging from
millisecond to years, and it is mostly in the LH state. During
transitions between the LH and HS states, both gradual and rapid
variations in the hardness ratio have been observed.

The LH state is characterized by a low flux in soft X-rays and strong hard X-ray
and soft $\gamma$-ray flux. Its spectral variability on the time scale of days
is then weak, while the $\sim$50 keV flux changes by a factor of a few (e.g.,
Gierli\'nski et al.\ 1997). The X-ray spectrum in the LH state is well described
by the sum of an intrinsic power law with the photon spectral index, $\Gamma\sim
1.5$--2, which is then cut off at energies $\ga 100$ keV, and a Compton
reflection (Magdziarz \& Zdziarski 1995) component. The intrinsic spectrum is
due to thermal Comptonization in a plasma with an electron temperature of
$kT \sim 100$ keV and a Thompson optical depth of $\tau\sim 1$
(Gierli\'nski et al.\ 1997), similarly to the case of other BHBs in the LH state (e.g., Zdziarski et al.\ 1998).

Recently, Zdziarski et al.\ (2002, hereafter Z02) presented an
analysis of {\it CGRO}/BATSE and {\it RXTE}/ASM observations of
Cyg X-1. One of their most interesting results in the X/$\gamma$
range is that most of the LH variability on time scales of hundred
of days and longer can be explained by varying the slope of the
overall spectrum with a pivot energy between 20 and 100 keV. The
pivoting can be due to either variable plasma temperature or
variable optical depth (Z02). Broad-band X/$\gamma$ observations
are crucial in providing constraints on the nature of the
variability, and, in particular, on the role of the soft photons
in cooling the X-ray emitting plasma and the role of Compton
reflection. Here, we present results from the IBIS instrument
(Ubertini et al.\ 2003) on board of \integral\/ (Winkler et al.\
2003).

\begin{figure}
  \centering
  \includegraphics[width=8.5cm]{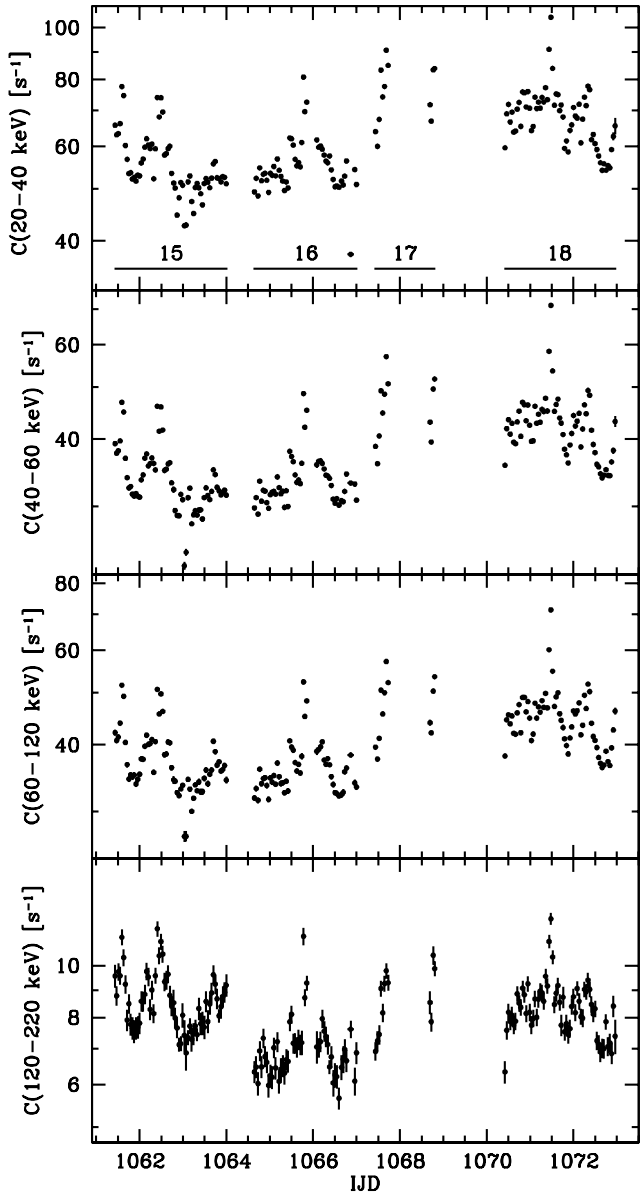}
 \caption{The count rates during the rev.\ 15--18 vs.\ the IJD ({\it INTEGRAL\/} Julian Day, starting from 2000
 January 1, 00.00 UT = JD 2451544.5, which corresponds to MJD = IJD + 51544). The extent of the data corresponding
 to a given \integral\/ revolution is marked in the top panel. The rates are shown in logarithmic scale with the
 same length per decade for each panel in order
to enable comparison of the variability in different channels. }
  \label{lc_cr}
\end{figure}

\section{Observation and data analysis}
\label{data}

Given the {\it INTEGRAL\/} launch date, Sun aspect constraints led
to Cyg X-1 being chosen as the first strong $\gamma$-ray target
for the initial calibration of the instruments during the
Performance Verification (PV) Phase. The observations of the
Cygnus region in general allowed for verification of the imaging
performance and the Point Spread Function of IBIS as well as its
cross-calibration with SPI and JEM-X. The PV phase consisted of a
staring, on-axis, observation lasting $\sim 850$ ks, and $\sim
1130$ ks of dithering observations to test the SPI performance.
Also, 68 ks were devoted to calibration of the JEM-X with
pointings of $1\degr$--$6\degr$ off axis. In addition,
contemporaneous observations of Cyg X-1 with {\it RXTE\/} were
performed (Pottschmidt et al.\ 2003).

The IBIS instrument (Ubertini et al.\ 2003) is a coded-mask
imaging telescope with a large field of view ($29\degr\times
29\degr$) based on two detector layers, ISGRI (Lebrun et al.\
2003) and PICsIT (Labanti et al.\ 2003) operating in the ranges 15
keV--1 MeV and 175 keV--10 MeV, respectively.  It provides fine
imaging ($12\arcmin$ FWHM) for source identification and good
spectral sensitivity in both continuum and broad lines over its
operative range. During the observations, IBIS was still
undergoing fine-tuning. Therefore, we studied only the staring,
on-axis, observations performed by IBIS in a stable configuration,
which correspond to revolutions 15 to 18, and use the consolidated
data reprocessed by the {\it INTEGRAL\/} Science Data Center
(Courvoisier et al.\ 2003). We use the offline scientific analysis
(OSA) software v.\ 1.1. Details and procedures on the IBIS
software developed by the IBIS Consortium are described in
Goldwurm et al.\ (2003). We have analyzed ISGRI single interaction
events for which arrival time, energy deposition and interaction
pixels are known. Most of the time, PICsIT was in its  standard
configuration that provide images at time interval of the order of
several minutes (1750--3600 s).

We produced the countrate, $C$, lightcurves during 2002 Nov.\
27--Dec.\ 8 binned by the pointings (science windows, hereafter
scws) in the ranges of 20--40, 40--60, 60--120, 120--220  and
220--350 keV. The typical duration of a scw is $\sim 2000$ s,
which allows us to study variability on the time scales ranging
from that time to days. The highest of the above energy channels
has not been used due to low statistical significance of its
single scw.

The lightcurves in the four energy bands are shown in Fig.\ \ref{lc_cr}. We see
that they are rather similar to each other. This is confirmed by calculating the
fractional rms variability (after subtracting the variability due to the
measurement errors), which is found to be $17\pm 1\%$, $18\pm 1\%$, $16\pm 1\%$,
$14\pm 1\%$ for channels with the increasing energy, respectively.

\begin{figure}[htbp]
  \centering
  \includegraphics[width=8.5 cm]{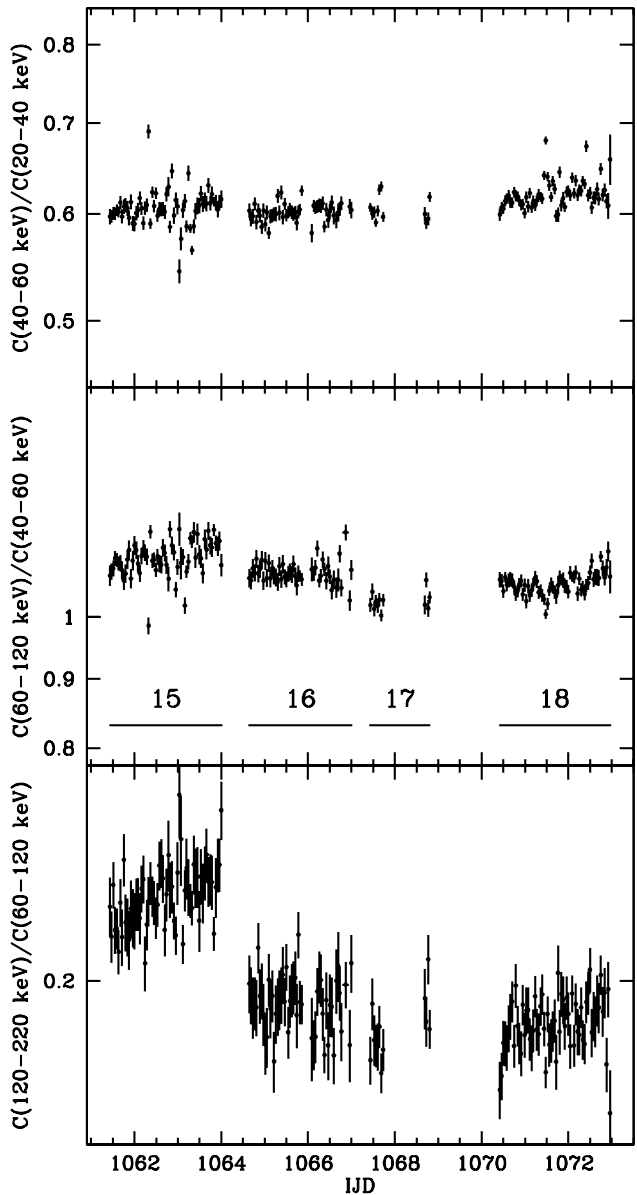}
  \caption{Hardness ratios for pairs of adjacent channels vs.\ time. The time interval of a given \integral\/ revolution
  is shown in the middle panel. The ratios are shown in logarithmic scale with the constant length per decade
  in order to enable direct comparison of the relative spectral variability. }
  \label{lc_hr}
\end{figure}

Fig.\ \ref{lc_hr} shows the corresponding evolution of the hardness ratio in
adjacent channels. As expected by the similarity of the lightcurves, variability
of the hardness ratios is much weaker, with the values of the fractional rms of
$3.4\pm 0.2\%$, $2.9\pm 0.2\%$, and $9.3\pm 0.5\%$ (with increasing energy).
Interestingly, the spectral variability is strongest at the highest observed
energies, $\ga 60$ keV. There is small, but statistically significant, increase
and decrease of the first and second hardness with time. Formally, the best
exponential fits give the e-folding times of $300\pm 20$ days and $220\pm 10$
days, respectively. On the other hand, there is a strong drop in the emission
$\ga 120$ keV from rev.\ 15 to 16 followed by a slow decrease during rev.\ 16,
and a slow increase during rev.\ 18. A time scale for significant changes of
that emission is $\sim 3$ days.

\section{Hardness-flux correlations}

Most of the information on the long-term characteristics of Cyg X-1 in hard
X-rays/soft $\gamma$-rays to date comes from the BATSE detector. In the LH
state, the 20--300 keV data show a positive correlation between the ratio of the
100--300 keV and 20--100 keV fluxes and the 100--300 keV flux. On the other
hand, the correlations of the opposite sign are seen in the range of the ASM
detector, 1.5--12 keV. Together, the data show the presence of a pivot in the
$\sim 20$--100 keV range (Z02). This is also supported by the fractional
variability as a function of the energy which has a minimum in the 20--100 keV
range. As pointed out in Z02, the pivoting variability appears only on very long
time scales, of the order of $\ga 100$ days. In addition, there is a second
variability pattern, in which the broad-band spectra move up and down with a
little change of the spectral shape. The presence of both patterns is also
confirmed by analysis of individual pointed observations (Z02, Gierli\'nski et
al.\ 1997).

The pivoting variability pattern is explained by Z02 as caused by a variable
luminosity in the soft seed photons irradiating the plasma. If the plasma is not
dominated by e$^\pm$ pairs, the variable irradiation results in the plasma
temperature adjusting itself to satisfy the energy balance. As a consequence,
softer spectra are expected to correspond to a lower electron temperature. This
appears to be confirmed by a \sax\/ observation in 1996 September (Frontera et
al.\ 2001), as pointed out by Z02. On the other hand, variations of the local
accretion rate in an a flow of a constant geometry may be responsible for the
variable amplitude with the constant spectral shape. In summary, Z02 found
strong changes in the hardness on long time scale, whereas variability was
dominated by changing the total luminosity on shorter time scales.

\begin{figure*}[htbp]
  \centering
  \includegraphics[width=11 cm]{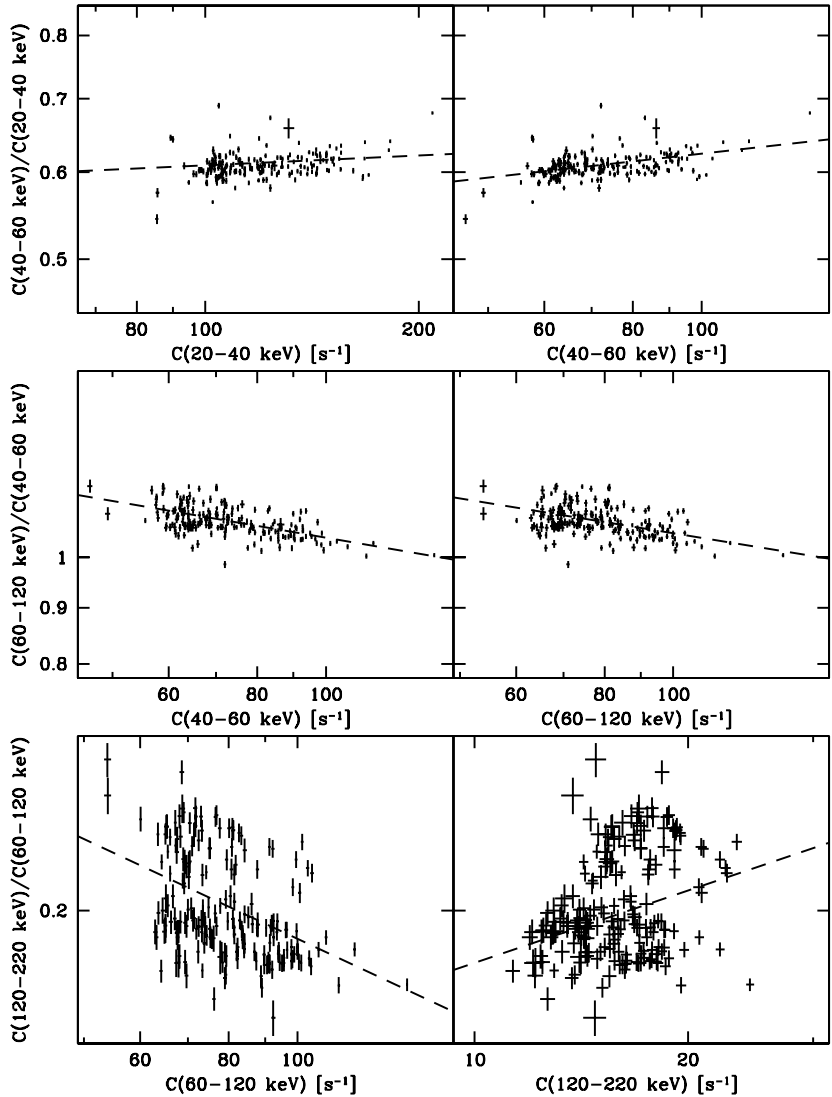}
  \caption{Hardness ratios for pairs of adjacent channels as functions of the count rates.
  The ratios and rates are shown in logarithmic scales with the same respective length per decade
  in each panel in order to enable direct comparison of the relative variability. The dashed lines
  correspond to best fit power-law dependencies. }
  \label{hr}
\end{figure*}

Following the above suggestions, we consider here correlations
between the spectral hardness and count rates. Our results are
shown in Fig.\ \ref{hr}, showing the three ratios (Fig.\
\ref{lc_hr}) each vs.\ their defining count rates. All the
hardnesses and count rates are correlated at a very high
statistical significance (see below). We find a new and unexpected
result that while the 40--120 keV hardness ratio decreases with
their defining count rates (top panels in Fig.\ \ref{hr}),
consistent with the pivoting variability of Z02, the 20--60 keV
hardness {\it increases\/} with the corresponding count rates.
(Note that such a behaviour could not be found by Z02, who used
only two rather broad BATSE channels, 20--100 keV and 100--300
keV.) The change of the sign of the hardness-flux correlation in
the 40--60 keV channel cannot be due to pivoting variability, as
hardness decreases with flux below a pivot and increases above it,
just opposite to our result.

The observed behaviour has to be due to a variable spectral curvature, with a
hump around $\sim 50$ keV appearing in the spectrum at high fluxes. Such a hump
is very likely to be due to Compton reflection, well known to be present in the
spectra of Cyg X-1 (e.g., Gierli\'nski et al.\ 1997; Gilfanov, Churazov \&
Revnivtsev 1999). A spectrum from reflection of an intrinsic BHB LH-state
continuum is very hard below its peak at several tens of keV and very soft above
it  (e.g., Magdziarz \& Zdziarski 1995). Thus, its addition hardens the total
spectrum below the peak and softens it at higher energies, entirely consistent
with our results. A quantitative study of this effect would require data at
energies $<20$ keV, not available to us at this time, as well as spectral fits
by each scw, which is beyond the scope of this Letter. However, we point out
that Compton reflection is known to be variable in Cyg X-1, and in particular it
correlates with the X-ray spectral index (Zdziarski, Lubi\'nski \& Smith 1999;
Gilfanov et al.\ 1999).

Going to higher energies, we also see that the hardness-flux correlation
continues to be negative up to $\sim 120$ keV, but it becomes again positive at
$\ga 120$ keV. This, on the other hand, is consistent with the findings of Z02
(see their fig.\ 6) and may indeed be due to spectral pivoting.

Fig.\ \ref{hr} also shows the best fit power-law dependencies, corresponding to
the count rate ratio $\propto C^\beta$. The values of the exponent are
$\beta=0.030\pm 0.004$, $0.073\pm 0.004$, $ -0.11\pm 0.01$, $-0.10\pm 0.01$,
$-0.30\pm 0.02$, and $0.22\pm 0.03$, for the six consecutive correlations. The
small relative uncertainties demonstrate the high significance of the
correlations. This is also confirmed by the Spearman rank correlation test,
which gives relatitvely high correlation coefficients of 0.29, 0.43, $-0.57$,
$-0.43$, $-0.42$, and 0.24. The corresponding probabilities of the absence of a
correlation are very low, with the highest value of $10^{-3}$ for the last
correlation and $\ll$ than that for all others.

\section{Spectral evolution}

We clearly see in Fig.\ \ref{lc_cr} a number of short flares and
dips (note that those are not absorption dips). To study in detail
evolution of the spectra, we have selected the scws corresponding
to the two of the flares and one of the dips. For reference, we
used stable periods containing 9 and 21 scws during the rev.\ 16
and 18, respectively. The log of the selected periods is given in
Table \ref{log}.

\begin{table}[htbp]
\centering
\caption{The log of the observations selected for the study of spectral evolution. }
\begin{tabular}{|l|c|c|c|c|} \hline \hline
Rev. & scw & type & IJD & exposure \\ \hline \hline 16 & 6--14 &
stable & 1064.6263--1064.9927 & 25269 s \\ \hline 16 & 26--27 &
peak & 1065.4406--1065.5220 & 5428 s\\ \hline 18 &  2--21 & stable
& 1070.4397--1071.2539 & 59816 s \\ \hline 18 & 26--28 & peak &
1071.4167--1071.5389  & 8491 s\\ \hline 18 & (2)22--24 & dip &
1072.5997--1072.7224  & 9557 s \\ \hline
\end{tabular}
\label{log}
\end{table}

At this stage of the IBIS programme a detailed and accurate
deconvolution matrix is still under finalisation and a spectral
analysis has to be carried out with caution. In this work we have
analysed the spectral data using a Crab-calibrated matrix with 64
energy channels. This matrix represents the Crab data well for
integration times not exceeding 100 ks and energies $\la 500$ keV.
We have added a 5\% systematic error to each PHA channel of the
spectra analyzed below. Spectral fitting was performed with the
standard XSPEC v.\ 11.2 tools.

\begin{figure}[htbp]
  \centering
  \includegraphics[width=8.5 cm]{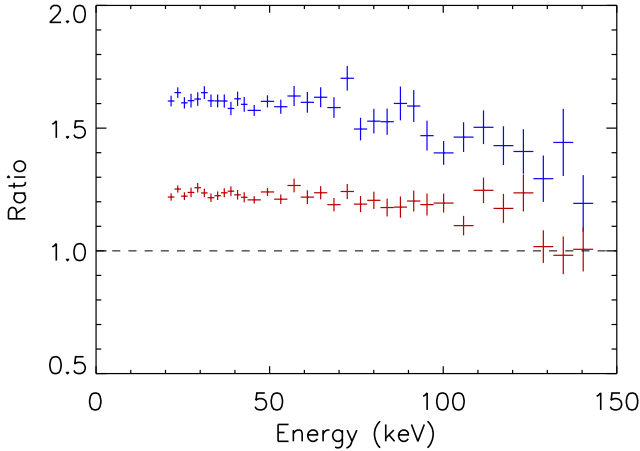}
\caption{Spectral evolution during the flare in rev.\ 18 as shown
by the ratios of the count rate spectra during scw 26 (red) and 27
(blue) to that from the scw 25 preceding the flare (see Table
\ref{log}).} \label{ratio1}
\end{figure}

To study the short term spectral evolution of Cyg X-1, we have selected three
spectra at the highest observed peak and just before it. This corresponds to the
day $\sim$1071.5 during rev.\ 18 and scw 25, 26 and 27 (the last one at the
maximum of the flare), see Fig.\ \ref{lc_cr} and Table \ref{log}. The three data
sets have been fitted to look for spectral evolution on the time scale of $\sim
3000$ s over a wide energy range. A good fit ($\chi_\nu^{2}\sim 0.9$--1.1) has
been obtained with a thermal Comptonization model (comptt of Hua \& Titarchuk
1995) at the electron temperature of $kT\sim 50$ keV, the optical depth of $\tau
\sim 0.9$, and the seed photon temperature constrained only to $\la 1$ keV.
Nevertheless, the model parameters are not well constrained due to the
relatively short exposure, which causes the photon statistics to be poor at high
energies, $\ga 150$ keV. To overcome this problem, we have tried a simpler
model, namely, a power law. This yielded the spectral indices of $\Gamma=2.04\pm
0.02$, $2.09\pm 0.02$, $2.14\pm 0.02$ for scw 25, 26, 27, respectively. Although
$\chi_\nu^{2}$ are always worse than those of the comptt fits, the above numbers
clearly show a softening of the spectrum with the increasing flux. Fig.\
\ref{ratio1} confirms this results by showing the ratios of the count rate
spectra at scw 26 and 27 to that at the scw 25. We see that the softening takes
place mostly at the highest energies, $\ga 90$ keV.

\begin{figure}[htbp]
  \centering
  \includegraphics[width=8.5 cm]{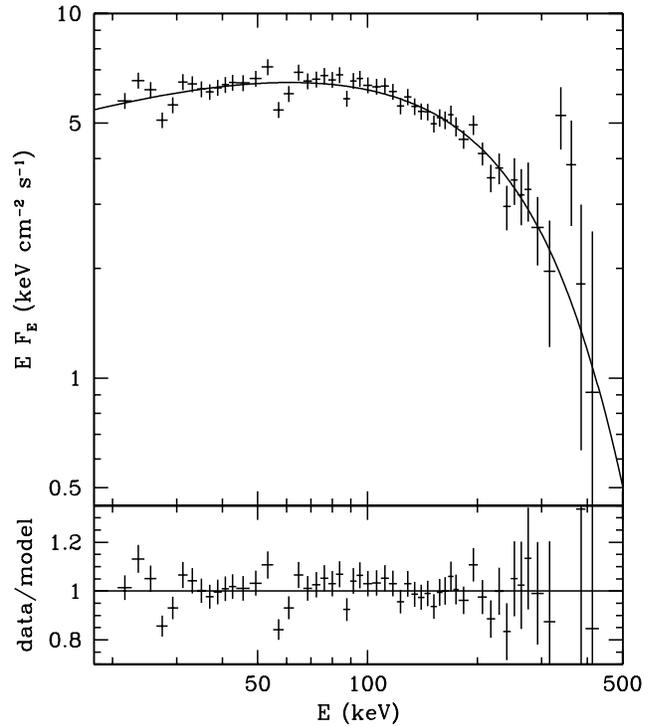}
  \caption{The average physical spectrum for scw 1--33 of rev.\ 18, fitted by a
  thermal Comptonization model. The bottom panel shows the fit residuals.}
\label{spectrum}
\end{figure}

Given the above problems with photon statistics, we have obtained
a spectrum from an uninterrupted period of 105 ks during rev.\ 18
(scws 1--33, IJD 1070.4019--1071.7310) during which the source did
not show significant spectral variation. A power law model does
not provide any good fit, and we used the comptt model, as above.
The obtained parameters are $kT = 64 \pm 9$ keV, $\tau=0.7\pm0.4$
and the seed photon temperature constrained to $\la 1$ keV, which
yields $\chi_\nu^2\simeq 1.5$ with 51 d.o.f.. For comparison, the
Crab fit with a power law yields $\Gamma = 2.17\pm 0.01$ at the 1
keV normalization of $11.5\pm 0.5$ cm$^{-2}$ s$^{-1}$ at
$\chi_\nu^{2}\simeq 1.3$ for 56 d.o.f. The average physical
spectrum is shown in Fig. \ref{spectrum}.

\begin{figure}[htbp]
  \centering
  \includegraphics[width=8.5 cm]{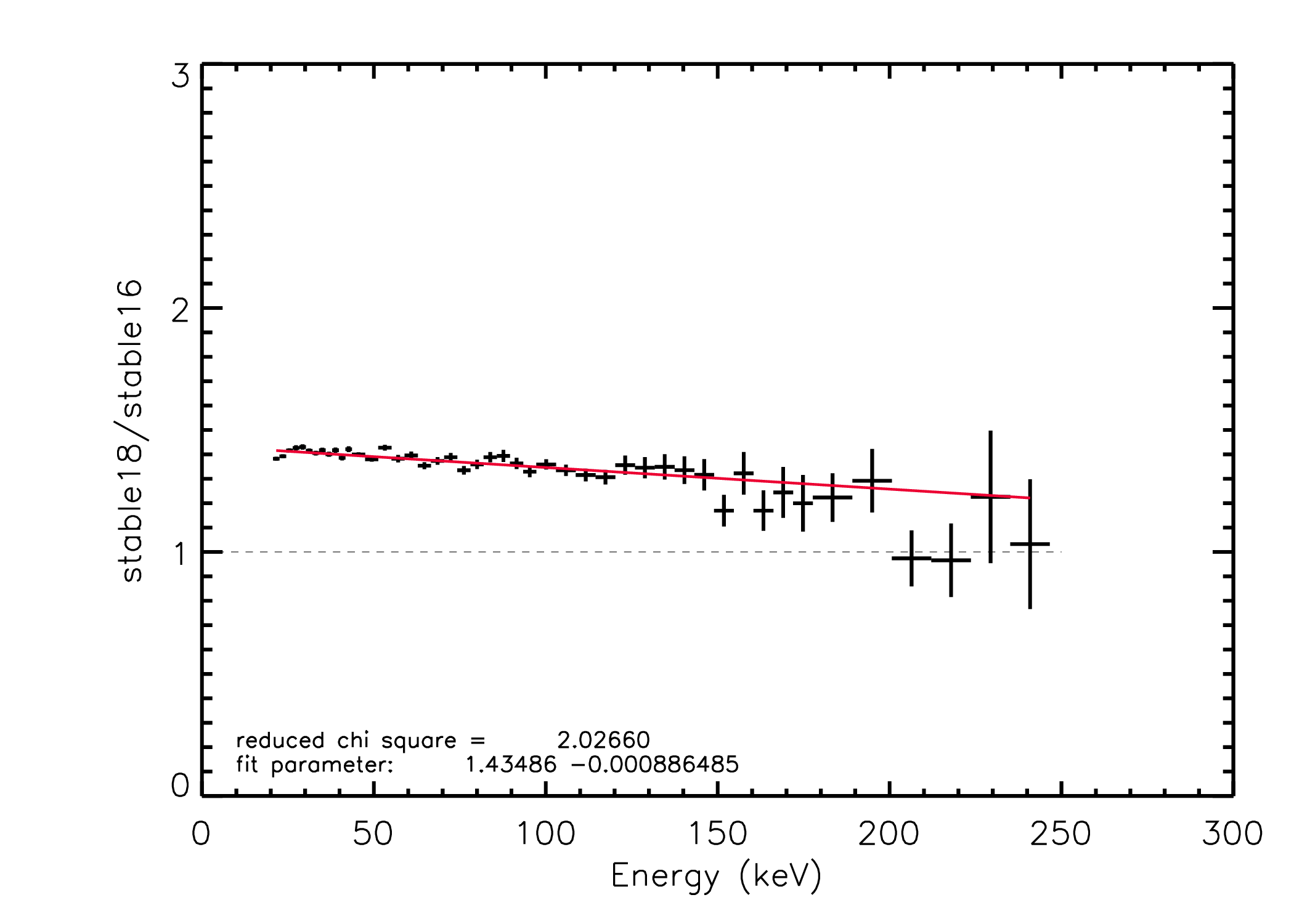}
  \includegraphics[width=8.5 cm]{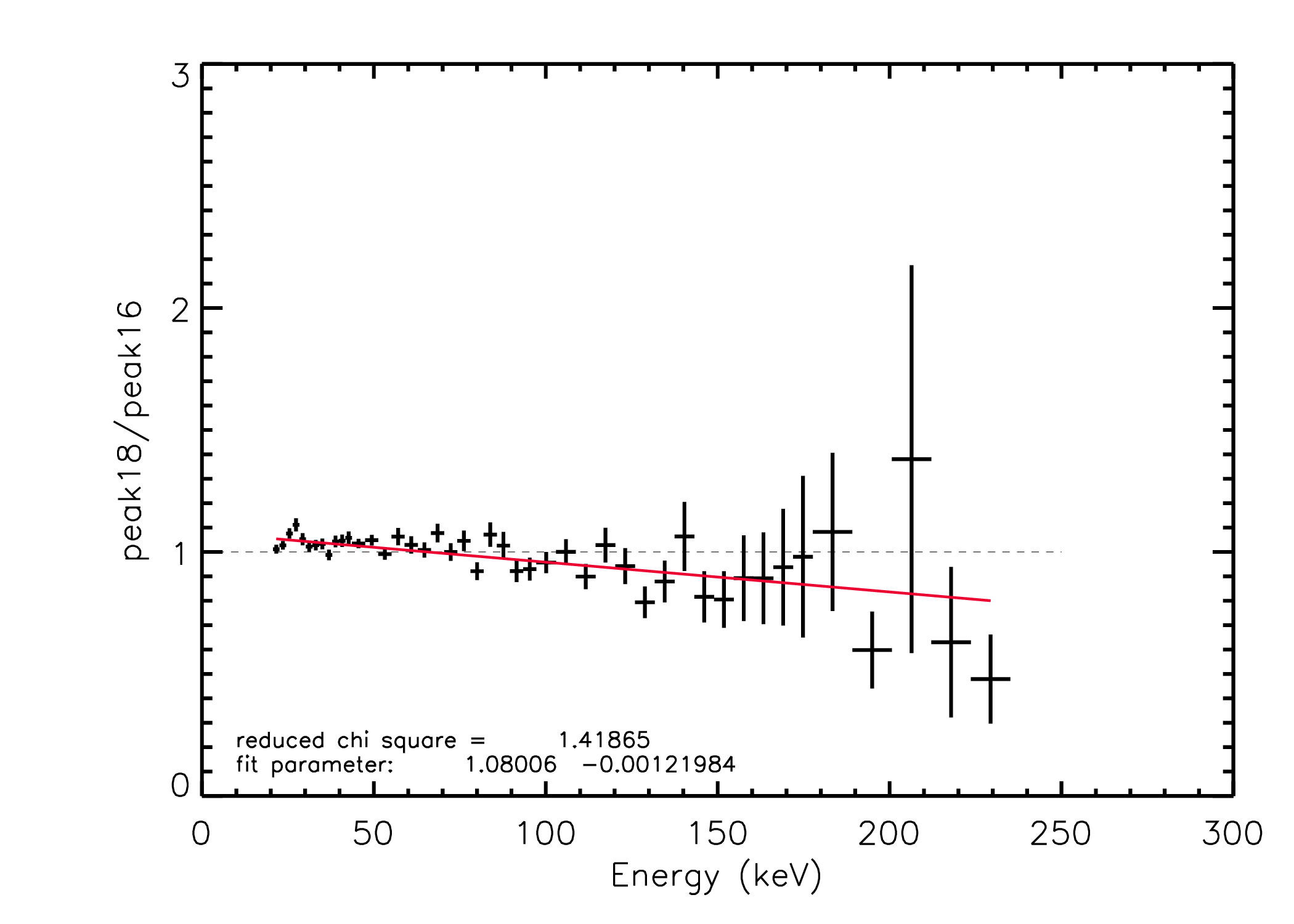}
  \includegraphics[width=8.5 cm]{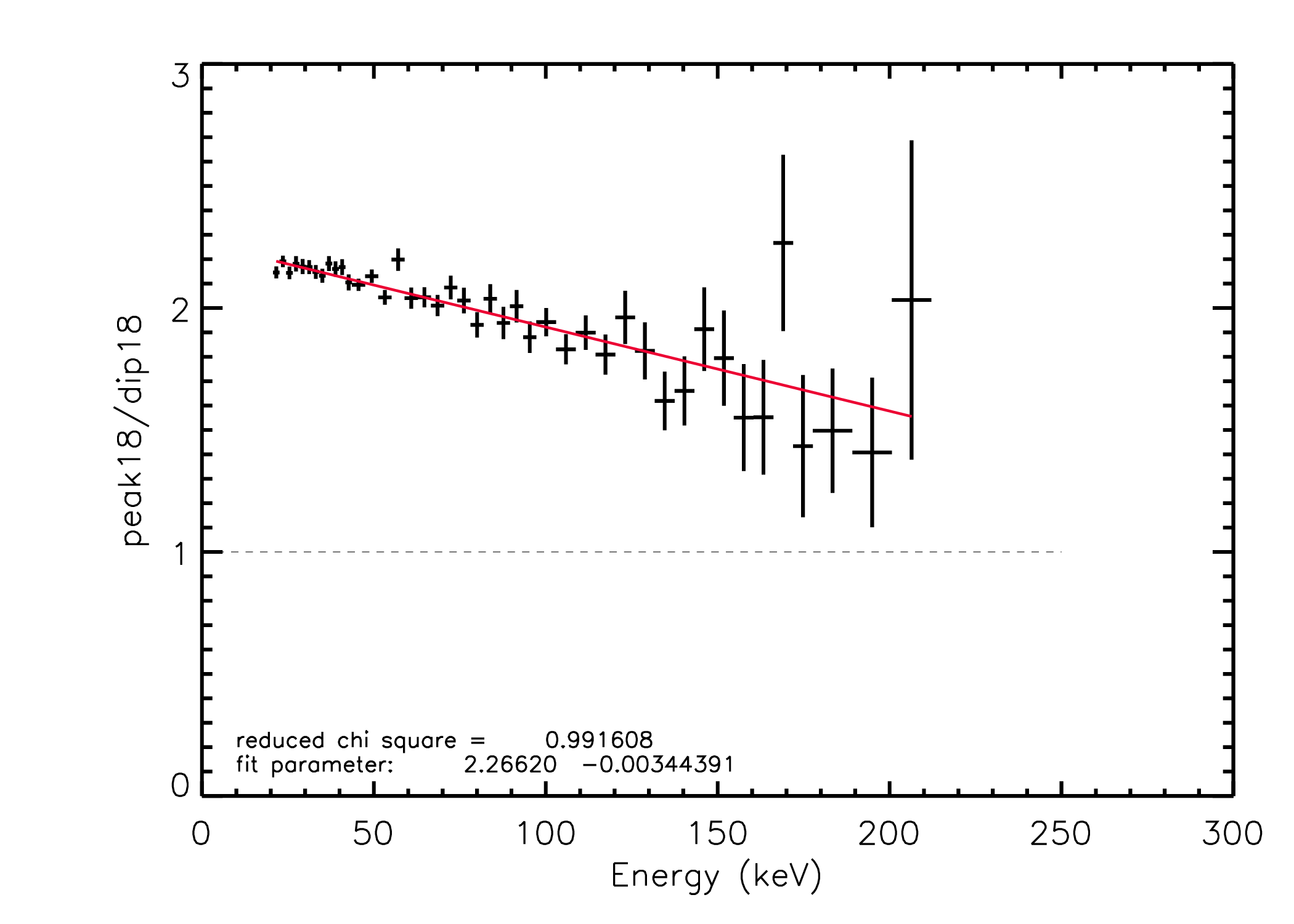}
  \caption{Count rate ratios between different spectra listed in Table 1. Top: The ratio between spectra
  during the stable period in rev.\ 18 to that in rev. 16. Middle: The ratio between the peak in rev.\ 18 to
  that in rev. 16. Bottom: The peak-to-dip ratio during rev.\ 18. }
 \label{ratio2}
\end{figure}

To further study correlations between the flux and the spectral hardness, we
have measured the count-rate ratios between the five spectra listed in Table
\ref{log}. This method allows us to eliminate possible uncertainties due to
spectral deconvolution. In Fig.\ \ref{ratio2}, we plot the ratio between two
stable periods, two peaks, and between a peak and a dip. The first two ratios
show no obvious trends, with the pairs of spectra just slightly differening in
the slope. On the other hand, we clearly see a significant softening with the
increasing flux in the peak-to-dip ratio (the bottom panel of Fig.\
\ref{ratio2}). However, none of the chosen spectra has allowed us to measure any
dependence of the high-energy cutoff (that would measure the plasma temperature)
on the flux.

\section{Conclusions}

We have shown that the long, almost uninterrupted, observation of Cyg X-1
performed during the PV phase with the IBIS broad-band imager represents an
invaluable database to deeply understand the nature of Cyg X-1 at high energies.
The presented IBIS data cover the energy range from 20 keV to a fraction of 1
MeV and a period of 12 days with interruptions of 7 hours every 3 days due to
the Earth radiation belt passage.

During that period, the flux varied by a factor of $\sim$3, including periods of
stable emission, flares and dips. We have first studied correlations between the
spectral hardness and the flux. Over the studied photon energy range, the
correlation was found to change the sign twice over the 20--220 keV; first from
positive to negative at $\sim$50 keV, and then back to positive at $\sim$120
keV. The former change appears to be due to the spectral curvature introduced by
variable Compton reflection. The latter may be due spectral pivoting.

The characteristic time scale for changing spectral hardness is
the shortest, $\sim$3 days, at the highest energies, $\ga 100$
keV. Most of the spectral variability is, in fact, taking place at
those energies. The dependence of the high-energy emission on the
overall spectral properties is relatively complex. In addition to
the hardness-flux correlations mentioned above, we have found a
spectral softening with the increasing flux during flares.

%Detailed spectral properties as measured by IBIS are presented in
%Laurent et al.\ (2003).

\begin{acknowledgements} We  wish to thanks all the IBIS team whose  effort made
the instrument  working. A particular thanks to G. La Rosa (IASF/Pa) and R. Much
(ISOC), without them the Performance Verification phase could not be successful
and to M. Federici for the long and valuable job to make the scientific analysis
possible at IASF/Roma. We also like to thank the ESOC Team for the continuous
support and kind hospitality. AAZ has been supported by KBN grants 5P03D00821,
2P03C00619p1,2, PBZ-054/P03/2001, and the Foundation for Polish Science. The
IBIS programme has been funded in part by the Italian Space Agency.
\end{acknowledgements}

\end{document}